\documentclass[electronic]{vgtc}             

\ifpdf
  \pdfoutput=1\relax                   
  \usepackage{graphicx}                
  \DeclareGraphicsExtensions{.pdf,.png,.jpg,.jpeg} 
\else
  \usepackage{graphicx}                
  \DeclareGraphicsExtensions{.eps}     
\fi%

\graphicspath{{figures/}{pictures/}{images/}{./}{png/}} 

\usepackage{microtype}                 
\PassOptionsToPackage{warn}{textcomp}  
\usepackage{textcomp}                  
\usepackage{mathptmx}                  
\usepackage{times}                     
\usepackage{cite}                      
\usepackage{tabu}                      
\usepackage{booktabs}                  

\usepackage{subfig}
\usepackage{caption}

\onlineid{0}

\vgtccategory{Research}

\vgtcinsertpkg

\title{Virtual Lenses as Embodied Tools for Immersive Analytics}

\author{Sven Kluge$^{\mbox{\scriptsize 1,}}$\thanks{e-mail: sven.kluge@uni-rostock.de} %
\and Stefan Gladisch$^{\mbox{\scriptsize 2}}$ %
\and Uwe Freiherr von Lukas$^{\mbox{\scriptsize 2}}$ %
\and Oliver Staadt$^{\mbox{\scriptsize 1}}$ %
\and Christian Tominski$^{\mbox{\scriptsize 1}}$ %
}
\affiliation{\scriptsize $^{\mbox{\scriptsize 1}}$Institute for Visual \& Analytic Computing, University of Rostock\\ $^{\mbox{\scriptsize 2}}$ Fraunhofer Institute for Computer Graphics Research IGD, Rostock}

\teaser{
  \centering
  \includegraphics[width=0.8\textwidth]{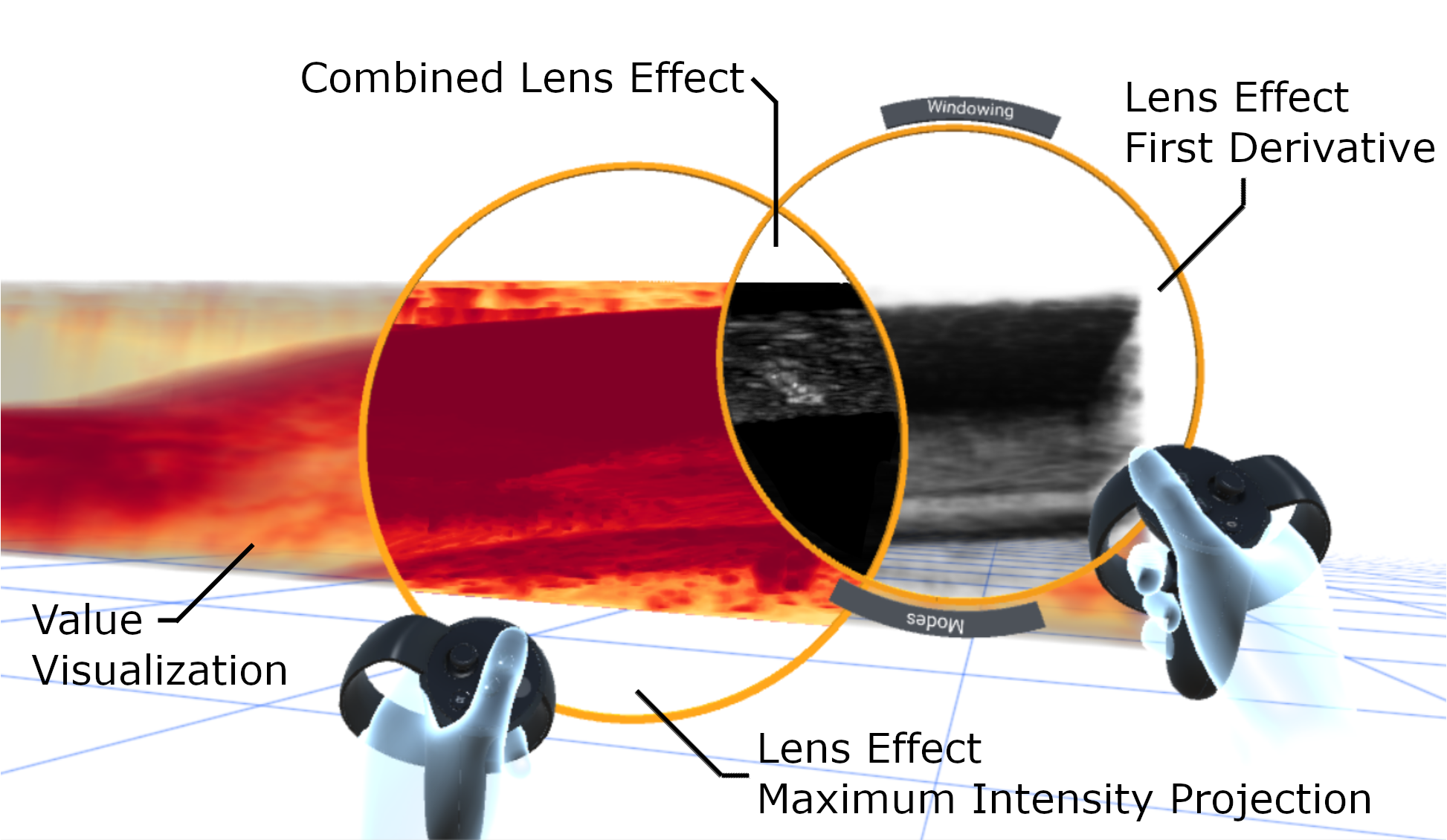}
  \caption{Combining the effect of two lenses to reveal a ship wreck hidden in a 3D sonar dataset.}
  \label{fig:lens-combination}
}

\abstract{
Interactive lenses are useful tools for supporting the analysis of data in different ways. Most existing lenses are designed for 2D visualization and are operated using standard mouse and keyboard interaction. On the other hand, research on virtual lenses for novel 3D immersive visualization environments is scarce. Our work aims to narrow this gap in the literature. We focus particularly on the interaction with lenses. Inspired by natural interaction with magnifying glasses in the real world, our lenses are designed as graspable tools that can be created and removed as needed, manipulated and parameterized depending on the task, and even combined to flexibly create new views on the data. We implemented our ideas in a system for the visual analysis of 3D sonar data. Informal user feedback from more than a hundred people suggests that the designed lens interaction is easy to use for the task of finding a hidden wreck in sonar data.
} 

\CCScatlist
{
  \CCScatTwelve{Human-centered computing}{Visu\-al\-iza\-tion}{Visu\-al\-iza\-tion techniques}{Treemaps};
  \CCScatTwelve{Human-centered computing}{Visu\-al\-iza\-tion}{Visualization design and evaluation methods}{}
}

\nocopyrightspace

\begin{document}


\firstsection{Introduction}

\maketitle



Visualization scientists have recognized the need for research into technologies that go beyond standard desktop environments with mouse and keyboard control~\cite{Lee2012}. In this context, the new emerging topic of \emph{immersive analytics} aims at exploring how data analysis can benefit from advances in virtual reality and augmented reality \cite{Marriott18Immersive}. The idea is to immerse users in their data to strengthen the interplay between human and machine in data analysis settings.

Typically, established visualization methods and interaction techniques cannot be applied in a straightforward fashion to new display environments and interaction modalities~\cite{Keefe13ReImagining}. Instead, it is necessary to rethink and adapt existing approaches to make them fit for the new technologies. In this work, we are particularly interested in the use of interactive lenses in the context of immersive analytics. Lenses are widely applied in visualization~\cite{Tominski2017}. Yet, most lens techniques are designed for classic 2D visual representations and standard mouse interaction. How interactive lenses can be utilized for immersive analytics remains an under-researched question. In particular the natural interaction with lenses in an immersive setting deserves more attention.

In this paper, we present a concrete interaction design for virtual 3D lenses for immersive analytics. Immersive 3D environments provide us the opportunity to enhance lenses by representing them as virtual and graspable tools. To make immersive lenses intuitive to operate in a direct and natural way, we employ the extended interaction capabilities of virtual reality (VR). Our interaction design focuses on utilizing the hands for creating, manipulating, parameterizing, and combining lenses. For lenses that are not within the user's reach, we create proxies that allow users to control lenses remotely. All interactions are aligned with the idea of \emph{fluid interaction} to promote flow and user immersion \cite{Elmqvist2011}.

Our techniques have been continuously tested in informal hands-on sessions by more than a hundred people over a period of two years. The acquired user feedback helped us to improve our design over several iterations. Overall, the feedback suggests that users can easily utilize our virtual lenses in immersive analytics environments.

\section{Related Work}

The work we present here is related to immersive analytics and interactive lenses. Both topics will be reviewed briefly in the following.

\subsection{Immersive Analytics}

Recent availability of affordable high-quality virtual reality (VR) and augmented reality (AR) devices, such as the HTC Vive or Microsoft Hololens, respectively, have opened up new opportunities for applying these technologies in the context of data analysis.

Research on immersive analytics aims at immersing users deeper in their data-driven analysis work~\cite{Marriott18Immersive}. Deeper immersion can potentially lead to higher user engagement, more conscious experiences, and smoother workflows in comparison to traditional desktop environments \cite{Rosenbaum11Immersive, chandler2015, hackathorn2016, Bach2016}. Especially when the primary dimensions of the data are spatial, immersive approaches can make the analysis more effective \cite{donalek2014}. Immersion can also lead to a higher degree of presence \cite{Slater1997, Cummings2015}. An important factor for this is the embodiment of the user, which can be reached by supporting  natural proprioception \cite{Mine1997}. In our case, this is supported by the head and hand tracking of the VR system. The definition of embodiment can also include every object in the virtual environment \cite{dourish2004action}. The ImAxes system presented by Cordeil et. al. \cite{Cordeil2017} explored the usage of embodied tools for multivariate data exploration in an immersive environment. These embodied tools behave like real physical objects and can be directly manipulated by the user. We pick up on this idea and create our virtual lenses as embodied tools.

\subsection{Interactive Lenses}

Interactive lenses are lightweight tools that provide a transient alternative representation of the data in a locally confined region of the display~\cite{Tominski2017}. Much of the value of lenses lies in the fact that they serve two purposes simultaneously: (1) they are the tool to select the part of the data or the virtual environment to be affected and (2) they represent the space where the alternative representation is shown.

Interactive lenses can be applied to different types of data and various analytical tasks. Tominski et al. provide a comprehensive survey of more than 50 lens techniques for visualization purposes~\cite{Tominski2017}. Still more lens techniques have been published since then, both for 2D and 3D visualizations. For example, Shao et al. allow users explore local regressions by moving lenses across a 2D scatter plot~\cite{Shao17RegressionLens}. Rocha et al. investigated how the lens concept can be adapted to support the exploration of data on 3D surfaces~\cite{Rocha18DecalLenses}. Their idea is to attach the lens to the 3D surface very much like a decal.

Previous work has also investigated more natural interaction with lenses. For example, Spindler et al. study tangible interaction for using lenses on and above a table-top surface display~\cite{spindler2010}. Others have applied lenses to AR~\cite{Brown2006, looser2007} and VR~\cite{Mota2016, Tong2017}. Most recently, Mota et al. explored the utilization of lenses for immersive analytics~\cite{Mota2018}. They present a concept based on the combination of a 3D spherical lens and a 3D surface decal lens. With their approach, the selection of the part of the virtual environment to be altered corresponds to a finite 3D volume of interest. In contrast to that, our lenses are designed as see-through tools with a (theoretically) unbounded selection volume. That said, the lens design presented in the following is a consequent next step towards exploring the benefits of interactive lenses in immersive data analysis settings.

\section{Natural Lens Tools for Immersive Analytics}

\begin{figure}[t]
	\centering
	\includegraphics[width=0.5\columnwidth]{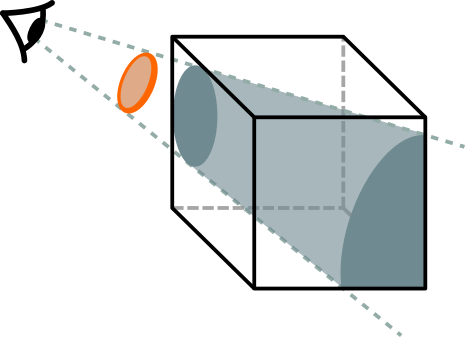}
	\includegraphics[width=0.4\columnwidth]{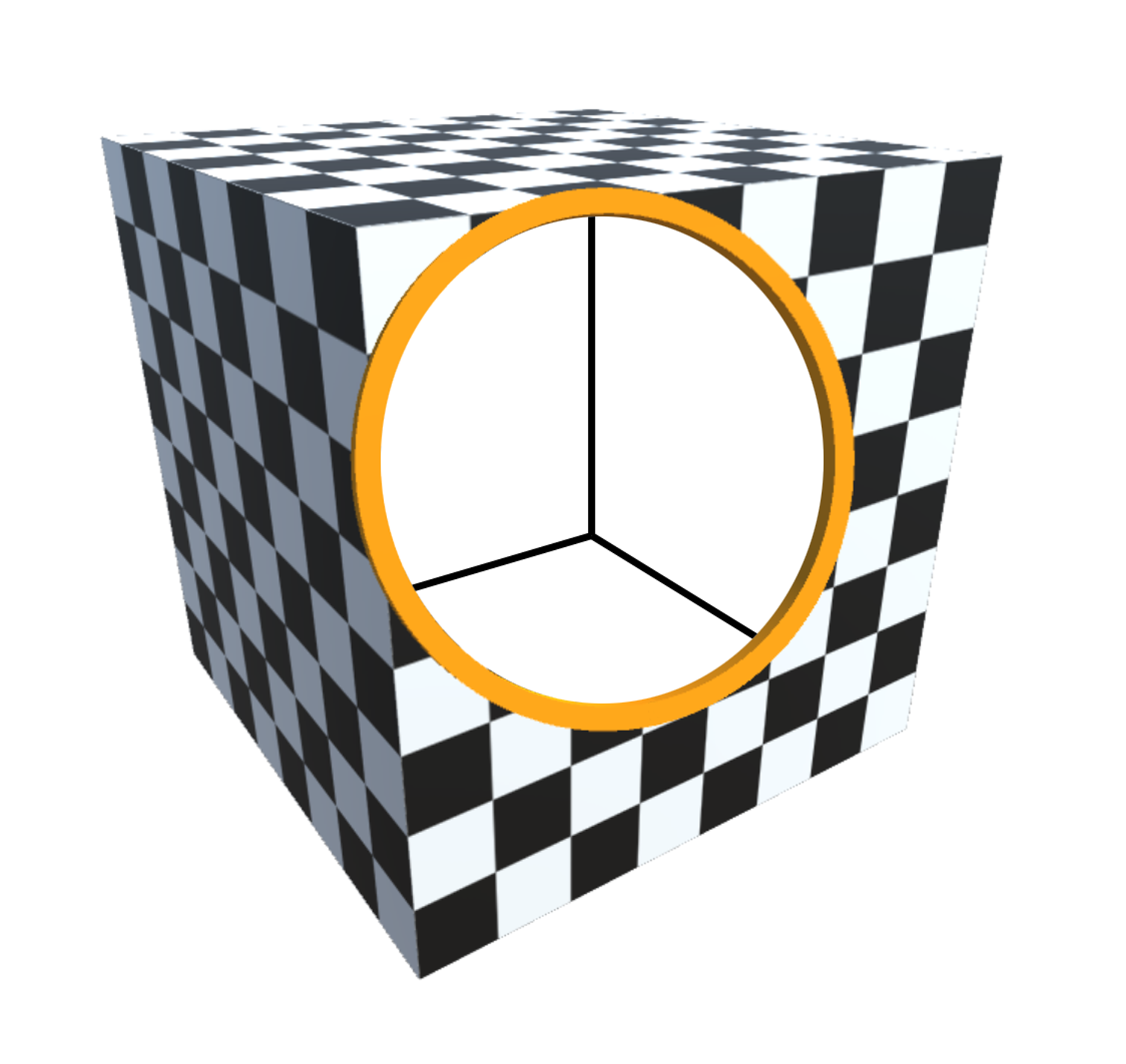}	
	\caption{Concept of virtual lenses.}
	\label{fig:lens-mockup}
\end{figure}

\begin{figure}[t]
	\includegraphics[width=\columnwidth]{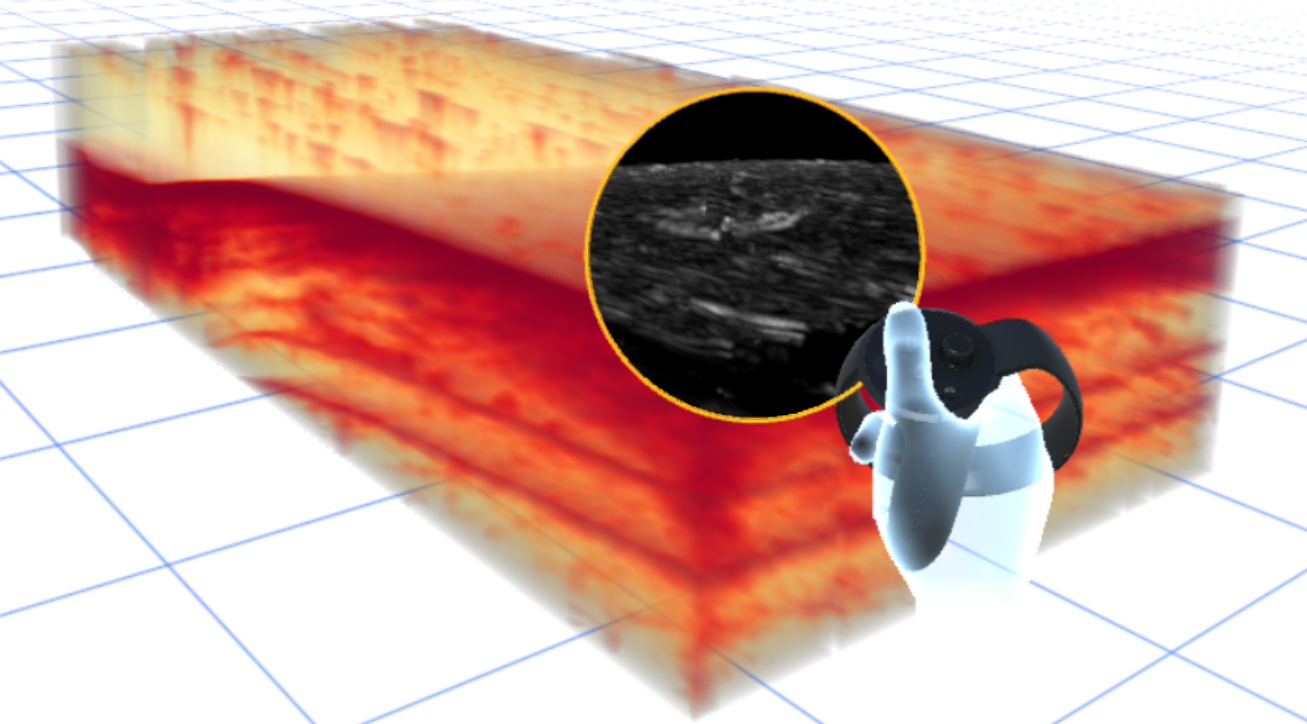}
	\caption{Using a virtual lens for looking into a 3D sonar dataset.}
	\label{fig:lens-display}
\end{figure}

Following the classic loupe metaphor, we decided to represent our virtual lenses as 3D discs that can be positioned and oriented freely in the environment. A conceptual view is depicted in \autoref{fig:lens-mockup}. The orange disc represents the lens through which the user can observe objects in the virtual environment. The lens interior will show an alternative view of the virtual world depending on the active lens effect. A lens ring serves as a clear delineation between the regular and the alternative view. As can be seen in a practical example in \autoref{fig:lens-display}, our virtual lenses look not that different from traditional interactive lenses. Yet, the way one can interact with them will be different.

\subsection{Direct Lens Interaction}

As mentioned before, our work concentrates on the question of how lenses can be interacted with utilizing the modalities of VR environments. There are three fundamental lens operations that need to be supported:

\begin{itemize}
	\item Create and remove: When a lens is needed, the user must be able to create and place one in the virtual environment. The inverse operation is to dismiss a lens when it is no longer needed.
	\item Manipulate and parameterize: Interactively exploring data with a lens primarily means manipulating the lens position, orientation, and size. This way, users can select the part of the display to be visualized in a different fashion. Parameters of the lens can further be adjusted to fine-tune the lens effect.
	\item Combine: Lenses naturally lend themselves to combining their individual effects by stacking them. Both, manual positioning of lenses and semi-automatic support for combining lenses are practical solutions.
\end{itemize}

We define naturalness and directness is the primary design goals, taking into account interaction costs~\cite{Lam08InteractionCosts}, especially physical motion costs and the time needed to execute an action. With this design rationale, we aim to reach a high degree of user immersion.

\paragraph{Lens creation and removal}

In the first place, we need a means to create lenses when they are needed for a particular task and to dismiss them once the task has been accomplished. To keep the focus of the user on the data and to avoid clutter in the virtual world, interaction tools such as lenses are stored in a semantically structured toolbox that allows the user to add or remove certain tools to or from the virtual world on demand.

\begin{figure}
	\includegraphics[width=\columnwidth]{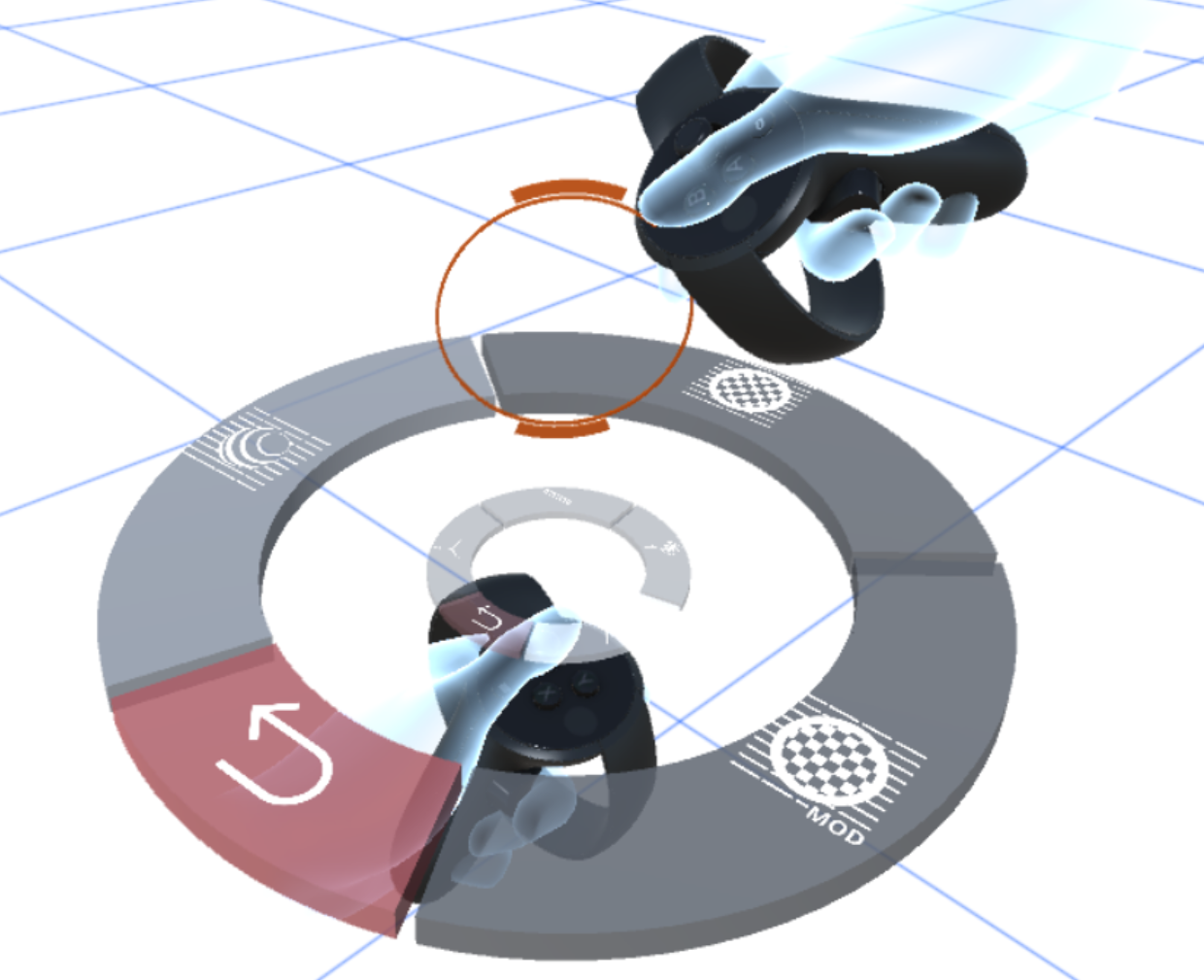}
	\caption{VR menu as a toolbox for lens creation and removal.}
	\label{fig:lens-menu}
	\vspace*{1em}
\end{figure}

The toolbox is designed as a multi-layered radial VR menu in which lenses of different type are available. The menu can be spawned by pressing a button at the controller in the user's non-dominant hand. A smooth animation brings up the menu from the virtual controller to a position above the user's virtual hand, as shown in \autoref{fig:lens-menu}. When the user moves the non-dominant hand, the menu moves along with it, which makes it easy to position the menu in a location that reduces occlusion of the visualization.

For each available lens, there is a corresponding section in the radial menu. Small icons in the menu symbolize the different lens effects. In order to create a lens, the controller in the dominant hand can be used to push down a section in the radial menu. The created lens then appears smoothly animated in the center of the menu. From there it can be grabbed directly with the user's dominant virtual hand (e.g., by using the hand trigger of an Oculus Touch Controller or a grab gesture with a Leap Motion device) and placed in the virtual environment. A lens will maintain its position and orientation unless the user deliberately manipulates it. When a lens is no longer needed, the user can simply grab the lens with the dominant hand, spawn the menu with the non-dominant hand, and put the lens back into the toolbox.

\paragraph{Lens manipulation and parameterization}

Easy manipulation of a lens is crucial, because it selects the part of the world for which an alternative view will be shown. For classic 2D lenses, the selection is dependent on the lens position and the lens size, which are typically controlled with the mouse. For 3D lenses in a virtual environment, the lens orientation and the user's head position are additional factors influencing the selection. Consequently, our virtual lenses are manipulated with the hands, while the head position represents an additional degree of freedom to adjust the perspective onto a lens, very much like in the real-world.

\begin{figure}
	\includegraphics[width=\columnwidth]{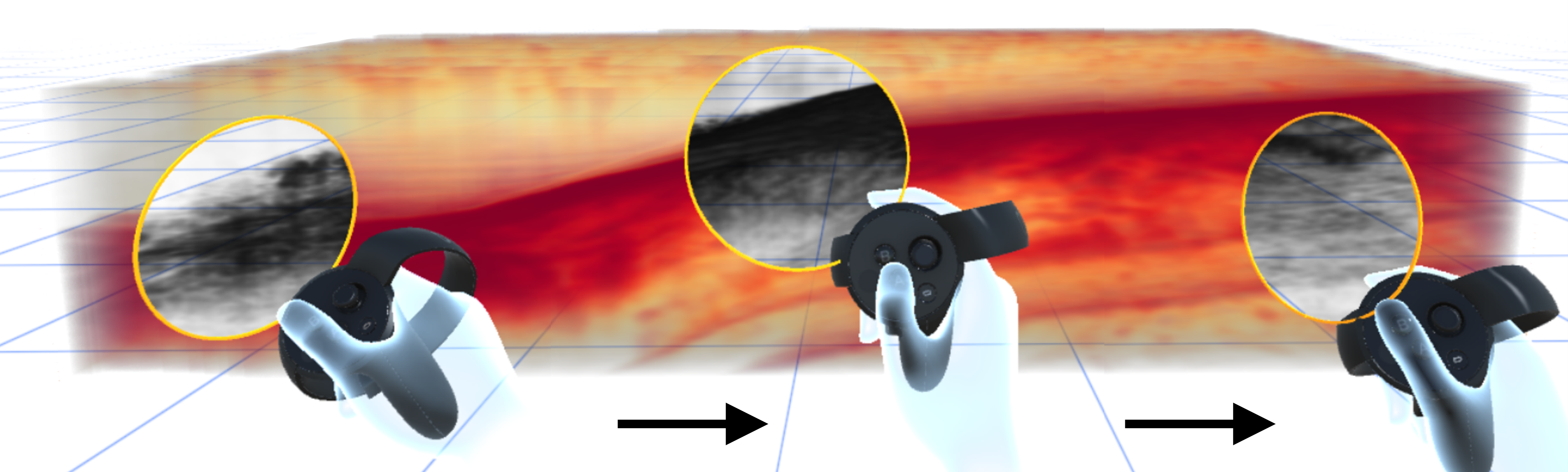}
	\caption{Positioning and orienting a lens with a virtual hand.}
	\label{fig:lens-translation}
\end{figure}

Positioning and orienting a lens are very natural in a virtual environment. As shown in \autoref{fig:lens-translation}, the user simply grabs the lens disc with either hand and manipulates it in six degrees of freedom as if dealing with a real-world object.

\begin{figure}
	\includegraphics[width=\columnwidth]{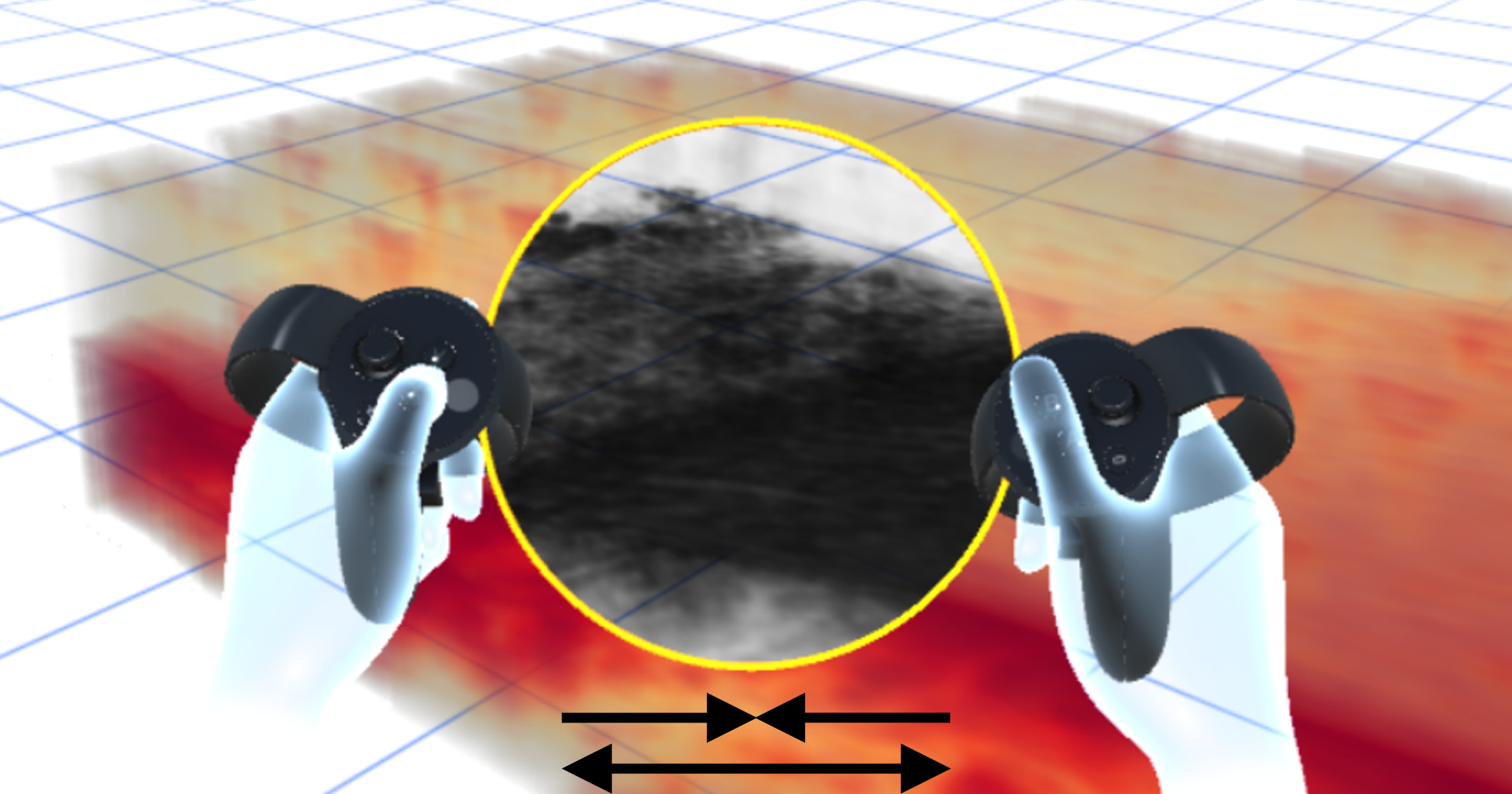}
	\caption{Resizing a lens with an adapted two-handed pinch gesture.}
	\label{fig:lens-scale}
\end{figure}

Real-world magnifying glasses are usually fixed in size, so there is no real-world inspiration for virtual resizing. Yet, there is a quasi-standard gesture for resizing: the pinch gesture. We adapted the two-finger pinch gesture to a two-handed resize gesture. As illustrated in \autoref{fig:lens-scale}, the lens is first grabbed with both hands. Then, while holding the lens ring, the hands can be moved in order adjust the lens size.

Manipulating a lens' position, orientation, and size allows users to select \emph{which} part of the world should be shown differently. Parameterizing a lens allows users to adjust \emph{how} the alternative representation should look like. We follow previous work and add user interface controls directly to the lens ring~\cite{Kister16SurfaceLens}. The controls enable the user to switch between a set of predefined lens effects and to adjust their parameters as needed. To minimize occlusion, the interface is kept compact and displayed only on demand. For a more direct selection of the lens effect, it is also possible to show different effects on the front face and the back face of the lens~\cite{spindler2010}. Switching between the two effects is then as simple as flipping the lens in the virtual world.

At any time, users can change their view through a lens by moving the head. On the one hand, this is very natural and intuitive. Yet, on the other hand, the head position is usually hard to fix in one place. Small movements of the head can substantially change the perspective towards a lens, and thus, the selection that is associated with it. This can make a detailed inspection of the visualized data difficult, because the view is constantly in flux. Solving this problem remains a task for future work, not only specifically for immersive lenses, but for immersive analytics in general.

\begin{figure*}
	\includegraphics[width=\textwidth]{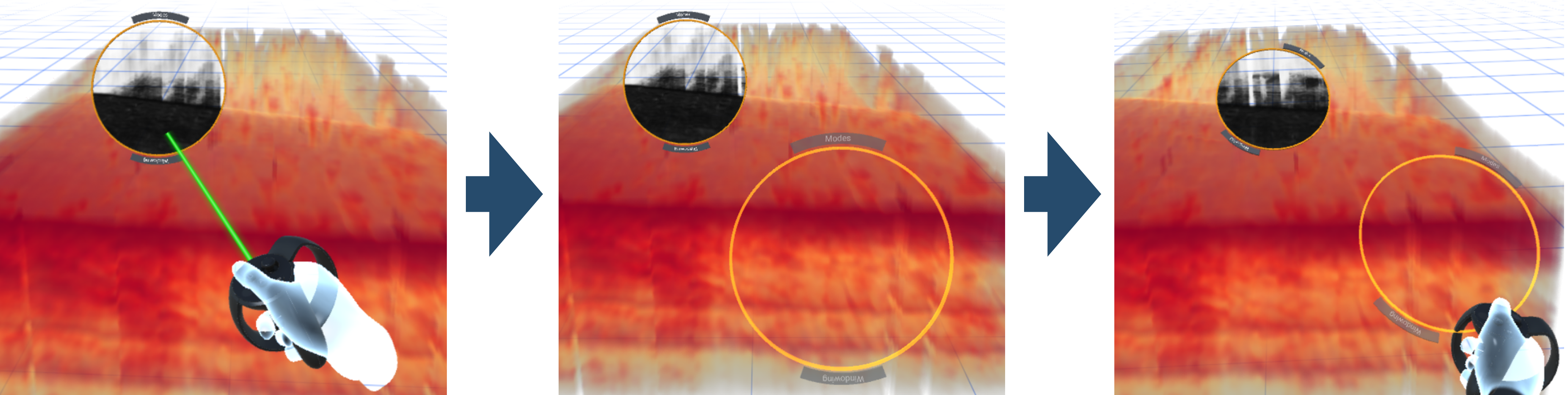}
	\caption{Far away lenses can be interacted with by selecting them with a raycast (left). This triggers the appearance of a semi-transparent proxy object (middle). Adjustments of the proxy will be applied to the remote lens (right).}
	\label{fig:lens-proxy-interaction}
\end{figure*}

\paragraph{Lens combination}


When multiple instances of lenses can co-exist in a virtual world, it is only natural to attempt to combine their individual effects. Based on the idea of Viega et al.~\cite{viega1996}, we enable users to combine the effects of lenses by grabbing and intersecting them much like combining camera filters in reality. Of course, this should be restricted to lenses whose combination leads to a meaningful and clear result. \autoref{fig:lens-combination} illustrates an example where a lens showing the data's first derivative is combined with a lens showing a maximum intensity projection. Overlapping lenses like this creates a composite lens effect that is temporary. It exists only as long as the overlap exists.

To permanently combine lens effects, the user has to create an almost maximal overlap of two lenses. Once the system detects the almost maximal overlap, the two lenses snap together and build a new single lens with the combined effect. The user interface elements of the individual lenses are rearranged at the new lens and an additional button is inserted for undoing the combination. This semi-automatic lens combination reduces interaction costs (one hand becomes free) and is particularly useful for combine several effects into one lens.

\subsection{Lens Interaction From Afar}

All interactions introduced so far assume that the user has the lens within arm's reach to control it directly with the hands. However, in a virtual world, this may not always be the case. The user may have moved away from the data in order to acquire an overview. The question now is how users can interact remotely with a lens that is not within their reach? Of course, one could walk up to the lens and grab it. However, this would be time-consuming and the user would lose the deliberately chosen point of view.

Addressing these issues, we integrated a technique for lens interaction from a distance which maintains the user's point of view and is still based on direct interaction. The idea is to interact with lens proxies, rather than with the remote lens. As illustrated in \autoref{fig:lens-proxy-interaction}, first, the user selects an out-of-reach lens via a raycast selection \cite{Bowman1997} (or through the toolbox). This triggers the animated appearance of a semi-transparent proxy of the lens in front of the user. The user can then interact with the proxy and all adjustments are automatically applied to the remote lens.

\section{Application Example and User Feedback}

The described interaction techniques have been implemented in a system to support the analysis of 3D sonar data. The system uses Oculus Rift devices and the Unity platform together with a modern graphics computer. The data contain measurements of the reflections of different materials in the water column and the seabed in the Baltic Sea. The original $80\mbox{m}\times40\mbox{m}\times4\mbox{m}$ volume with about 144M data points has been reduced to a $512\times256\times256$ resolution to achieve the high frame-rates required for VR.

For the collectors of the sonar data, it is of interest to examine local areas with high value changes (e.g., for analyzing structural anomalies) in the context of the global distribution of reflection values (e.g., for detecting material layers in which an anomaly is located). To support these tasks, several alternative visual representations of the data are necessary, which provides a nice opportunity to apply our lens techniques.

The base visualization is a direct volume rendering of the data. Users can walk around in the virtual environment and look at the data from different perspectives. Lenses can be utilized to look at different representations in a focus+context fashion. \autoref{fig:lens-combination} shows two virtual lenses visualizing the maximum intensity projection and the first derivative. In their overlap, both effects are combined. Only this combined lens effect reveals a buried wooden wreck in a material layer $0.5\mbox{m}$ below the seabed (bright structure inside lens overlap).

Using this application and the task of revealing the wooden wreck, we continuously tested our design in informal hands-on sessions. Over two years of development, we have acquired feedback from more than a hundred people from various backgrounds and domains (e.g., laypersons with no experience in VR or computer science, experienced researchers in the field of VR, and experts from the maritime domain). The sessions were typically unstructured public and internal lab demonstrations.

During early sessions, we could observe many users trying actions and expecting reactions resembling their real-world experiences. These observations inspired much of the design of our techniques. An example for this is the combination of lenses by overlapping. If our system did not support this operation, user expectations would not be met and the immersion would break. From the user feedback we also learned that multi-sensory feedback is important for a deep immersion. Therefore, we added haptic and acoustic feedback in addition to visual feedback.

The people with no experience in VR needed a short time to familiarize themselves with the environment and the interaction modalities. Once acquainted, they were rather enthusiastic about working with the lenses to track down the hidden wreck. Some of the people with VR experience had a particular problem with how the grabbing was implemented. They were used to commercial VR games where grabbing is typically done with the middle finger, and not with the index finger as in our system. This pointed us to add an option for switching the grabbing gesture between the two modes.

More abstract interactions such as using the radial menu for creating new lenses seem to be more prone to errors. For example, several users opened the radial menu, but then forgot about it and just by moving their hands accidentally pressed buttons of the menu.

Despite these minor issues, the overall feedback for the developed lenses was very positive. This suggests to us that our interaction techniques for virtual lenses can be useful in the context of immersive analytics. Yet, controlled studies are needed to actually confirm the benefits of naturally-inspired lens interaction and to compare different lens designs.

\section{Conclusion and Future Work}

We presented novel interactive lenses as virtual graspable tools for immersive analytics. We devised interaction techniques for working with lenses in a natural way, both directly and remotely. Our techniques have been implemented for the Oculus Rift and applied to support the analysis of a sonar volume dataset. Informal user feedback suggests that working with virtual lenses in VR can be intuitive and easy.

A particular concern to be addressed in future work is to decouple the lens manipulation from the (involuntary) head movement of the user. However, this is a delicate issue because the measures taken (e.g., Kalman filtering) might contradict with naturalness and hence might harm immersion. For further future work, we can imagine to extend our concepts to other data (e.g., lenses for graph visualization in VR) and to collaborative data analysis settings.


\bibliographystyle{abbrv-doi-hyperref}

\bibliography{template}
\end{document}